\documentclass[journal]{IEEEtran}

\usepackage{Style/myStyleIEEE}

\usepackage[margin=0.65in]{geometry}
\usepackage{booktabs}
\usepackage{tabularx}
\usepackage{amsmath}
\usepackage{ushort}
\usepackage{threeparttable}
\usepackage{makecell}
\usepackage{lettrine}
\usepackage{dblfloatfix}

\IEEEoverridecommandlockouts

\begin{document}

\title{\texorpdfstring{$\mathcal{H}_\infty$}{H-infinity} Grid-Forming Control in DC Microgrids}

\renewcommand{\theenumi}{\alph{enumi}}

\newcommand{\jovan}[1]{\textcolor{magenta}{$\xrightarrow[]{\text{J}}$ #1}}
\newcommand{\ognjen}[1]{\textcolor{pPurple}{$\xrightarrow[]{\text{O}}$ #1}}

\author{
    Jovan~Krajacic,~\IEEEmembership{Student~Member,~IEEE,}
    Ognjen~Stanojev,~\IEEEmembership{Member,~IEEE,}
    \\
    Mario~Schweizer,~\IEEEmembership{Senior~Member,~IEEE,}
    and~Gabriela~Hug,~\IEEEmembership{Senior~Member,~IEEE,}
    \thanks{J. Krajacic and G. Hug are with the Power Systems Laboratory, ETH Zurich, Switzerland. Emails: \{jkrajacic, ghug\}@ethz.ch. O. Stanojev and M. Schweizer are with the ABB Corporate Research Center, Baden-Dättwil, Switzerland. Emails: \{ognjen.stanojev, mario.schweizer\}@ch.abb.com}
    \thanks{\textit{Corresponding Author:} Ognjen Stanojev, ognjen.stanojev@ch.abb.com}
}

\IEEEoverridecommandlockouts

\maketitle
\IEEEpeerreviewmaketitle

\begin{abstract}
The growing share of DC-based generation, storage and loads has increased interest in DC microgrids. As a result, DC-bus voltage regulation and stability have become active research topics, with considerable effort directed at improving the transient performance of conventional droop control. Although numerous variants of droop algorithms have been proposed, many rely on increasingly complex control structures with multiple tuning parameters and measurements, thereby increasing the commissioning burden. Moreover, stable interaction with the surrounding microgrid is not guaranteed by design. To address these issues, this paper develops a passivity-constrained $\mathcal{H}_\infty$ synthesis method for the DC-bus voltage controller. The proposed $\mathcal{H}_\infty$ controller emulates the desired grid-forming behavior of the converter by minimizing the $\mathcal{H}_\infty$ error norm between device-level performance indices and their desired values, while additionally enforcing passivity during the optimization. The novel $\mathcal{H}_\infty$ control scheme is benchmarked in the frequency domain against representative droop-based and virtual-DC-machine algorithms from the literature, while experiments on a four-converter DC microgrid corroborate the frequency-domain analysis and validate the practical applicability of the algorithm.
\end{abstract}

\begin{IEEEkeywords}
DC microgrids, droop control, grid-forming, \texorpdfstring{$\mathcal{H}_\infty$}{H-infinity} synthesis, output impedance, passivity, voltage control
\end{IEEEkeywords}

\section{Introduction} \label{sec:intro}

The rapid growth of distributed energy resources and energy storage systems has increased interest in DC microgrids, since many of these resources operate inherently on DC or are interfaced through DC conversion stages~\cite{Jithin2023}. By eliminating unnecessary AC/DC and DC/AC conversion stages, DC microgrids can reduce conversion losses, converter count, and overall system cost~\cite{GEVernova2024, Justo2013}. Moreover, unlike AC microgrids, DC microgrids do not require reactive power management, frequency regulation, or AC synchronization, thus simplifying the overall control architecture~\cite{Dragicevic2016}. In DC microgrids, the primary grid-forming objective is DC-bus voltage regulation, which plays a role analogous to frequency regulation in AC~systems~\cite{Krajacic2026b}.

To regulate DC-bus voltage within prescribed limits while enabling decentralized load sharing, droop control has become a widely adopted solution for parallel source converters, as it allows coordination without communication among units~\cite{Austermann2024, currentos2025}. In this context, two droop-control variants are commonly considered in DC microgrids literature~\cite{Gao2017, Wang2018, Jin2017a}: voltage--current (V--I) droop and current--voltage (I--V) droop. In V--I droop, the measured output current is used to generate the converter voltage reference, whereas in I--V droop, the measured output voltage is used to compute the current reference. More generally, these two signal mappings define the V--I and I--V control architectures, with a proportional droop gain as their conventional realization. The former employs a dual-loop control implementation and therefore requires an additional current measurement, whereas the latter uses a single-loop implementation, thus reducing commissioning effort since only the DC-bus voltage measurement is needed. Nevertheless, neither structure is uniformly superior, and both variants have been extensively studied with the aim of improving their transient performance~\cite{Dragicevic2016a}.

Recent enhancements to conventional droop control can be broadly grouped into three categories: virtual-inertia emulation~\cite{Jin2018, Zhu2020, Lin2021}, machine-analogous control~\cite{Samanta2018, Tan2016, Na2022}, and output-impedance shaping~\cite{Jin2017b, Liu2020, Deng2024}. The first augments the droop loop with virtual energy-storage dynamics to enhance the inertial response beyond that provided by the physical DC-link capacitor, whereas the second, inspired by the virtual synchronous machine concept for AC-grid converters~\cite{DArco2015}, reproduces the electromechanical behavior of a DC machine to the same end. Both introduce additional tuning parameters and remain constrained by the converter control bandwidth. Furthermore, machine-analogous control requires a voltage-restoration controller constrained by the emulated machine dynamics. In contrast, output-impedance shaping modifies the converter dynamics directly to obtain resistive-capacitive impedance characteristics that improve both power sharing and transient disturbance rejection. Although effective, output-impedance shaping techniques rely on a first-order droop transfer function imposed by the adopted control structure rather than being freely selectable, which could limit impedance matching in certain cases. Moreover, the physical interaction of the converter with the surrounding network remains nontrivial due to the lack of guaranteed output-impedance passivity~\cite{Lazarevic2025}.

These limitations motivate a design procedure in which the desired converter behavior is specified rather than approached through iterative parameter tuning. For this purpose, $\mathcal{H}_\infty$ synthesis is well suited~\cite{Zhou1996}, since competing design objectives are expressed as frequency-domain weighting functions. The synthesis then yields a controller that minimizes the worst-case gain from the disturbance inputs to the weighted performance outputs over the entire frequency range. Moreover, structured and fixed-order formulations~\cite{Apkarian2006, Apkarian2014} allow the controller complexity to be fixed a priori, ensuring the resulting scheme remains implementable and interpretable, while additional stability requirements can be imposed as constraints during synthesis rather than verified a posteriori. In AC systems, these features have motivated extensive use of $\mathcal{H}_\infty$ synthesis for grid-following~\cite{Yang2011, Kammer2019, Haberle2022}, grid-forming~\cite{Chen2022}, and arbitrary~\cite{Huang2020} converter control strategies. In DC microgrids, by contrast, its use has remained sparse.

Most reported $\mathcal{H}_\infty$ designs in DC microgrids address the dual-active-bridge converter~\cite{Ren2021, Liu2024, Liu2025, Liu2026}, in which mixed-sensitivity weightings tune the loop that controls the phase-shift ratio to handle parameter uncertainty and input- and output-voltage disturbances, rather than to shape the behavior at the converter terminals. Among the few works that address the overall control design of DC microgrid converters using the $\mathcal{H}_\infty$ norm, \cite{Tripathy2025} synthesizes a robust voltage controller for a point-of-load buck converter through the $\mathcal{H}_\infty$ loop-shaping design procedure, combined with a virtual capacitor and conductance that emulate inertia. The synthesis, however, concerns a load-side unit and targets robust stabilization under parametric load uncertainty, thus neglecting the output impedance as a design variable, and the results remain limited to simulation. Conversely, \cite{Serrano2020} focuses on source-side converters, applying non-smooth $\mathcal{H}_\infty$ optimization to tune a fixed-structure droop controller such that the shaped output impedance allows the converter output capacitance to be reduced, and validates the design experimentally on buck- and boost-type half-bridge converters. Although effective, the method is formulated for the V--I control architecture, and thus inherits the need for the additional current measurement and the dual-loop implementation discussed above. Moreover, it treats the output capacitance as a free design variable, whereas in practice this quantity is typically fixed in advance by the converter configuration and its ratings. The relevant question is therefore not how to choose the capacitance, but what terminal behavior can be achieved with a given capacitance. Finally, since only magnitude requirements are imposed on the output impedance, the passivity of the resulting converter behavior is left unaddressed.

Motivated by the aforementioned research gaps, this paper proposes a passivity-constrained $\mathcal{H}_\infty$ synthesis method for an I--V type DC-bus voltage controller. The controller directly shapes the converter's output impedance and current-forming behavior toward their desired device-level profiles using the device-level indices established in our previous work~\cite{Krajacic2026a}, while enforcing passivity of the output impedance as a synthesis constraint. As a result, the controller achieves the desired grid-forming behavior of the converter, defined by the output-impedance index and the current-forming index. Furthermore, a sufficient passivity-based condition for stable interconnection is incorporated directly into the synthesis.

The proposed control algorithm utilizes the simple I--V control architecture and thus requires only the DC-bus voltage measurement to produce the current reference for the current controller, eliminating the need for additional measurements, such as output current or capacitor current feedback. The novelty therefore lies in the synthesis formulation through which a low-order dynamic controller is obtained to jointly track the desired terminal behavior while satisfying the passivity constraint. Furthermore, the underlying $\mathcal{H}_\infty$ tuning methodology can be readily extended to other control schemes and converter topologies. This contrasts with virtual-DC-machine-based approaches, which typically rely on nonlinear control structures and several machine-inspired parameters.

The novel I--V $\mathcal{H}_\infty$ scheme is analyzed in the frequency domain, providing parameter-tuning guidelines and characterizing the resulting output impedance. Furthermore, the control scheme is benchmarked against representative control algorithms from the recent DC microgrid literature using device-level performance indices \cite{Krajacic2026a}. Finally, the algorithm is experimentally validated on a DC microgrid setup, demonstrating its practical feasibility and low commissioning effort.

The remainder of this paper is organized as follows. Section~\ref{sec:GFM_control} introduces the concept of grid-forming converter behavior in DC microgrids, which is realized in this paper through the novel passivity-constrained $\mathcal{H}_\infty$ control design presented in Section~\ref{sec:Hinf_control}. The resulting I--V $\mathcal{H}_\infty$ controller is analyzed and benchmarked in the frequency domain against representative droop-based and virtual DC machine algorithms from the literature in Section~\ref{sec:results_freq}. Finally, Section~\ref{sec:results_exp} experimentally validates the I--V $\mathcal{H}_\infty$ controller on a DC microgrid setup, and Section~\ref{sec:concl} concludes the paper.

\section{Grid-Forming Control in DC Microgrids}\label{sec:GFM_control}

In this section, we establish the preliminaries needed to formulate the desired grid-forming behavior as a $\mathcal{H}_\infty$ synthesis objective. First, we introduce the proposed control architecture, in which a dynamic controller replaces the static droop gain of the I--V droop scheme to achieve the desired grid-forming behavior established in our previous work~\cite{Krajacic2026a}. Then, we derive the small-signal converter representation and the equivalent circuit from which the output impedance follows. On this basis, we summarize the device-level indices that quantify the voltage- and current-forming capability of the converter, and finally define the desired behavior that these indices should follow. Together, these elements constitute the plant model, performance measures, and reference behavior on which the $\mathcal{H}_\infty$ synthesis of Section~\ref{sec:Hinf_control} is built.

\subsection{Proposed Control Scheme}
\begin{figure}[t!]
    \centering
    \includegraphics[]{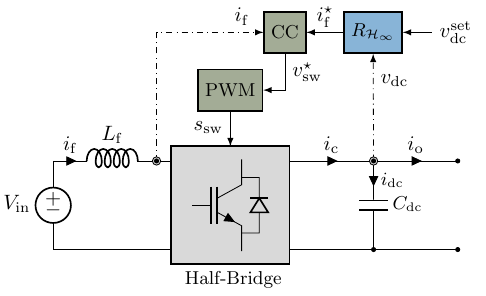}
    \caption{Proposed I--V $\mathcal{H}_\infty$ control structure in case of a bidirectional boost DC/DC converter, with the $R_{\mathcal{H}_\infty}$ controller highlighted in blue and the current controller and PWM in green.}
    \label{fig:Hinf_control}
\end{figure}

The proposed I--V $\mathcal{H}_\infty$ algorithm builds on the conventional I--V droop architecture, in which the output-voltage deviation $v_\mathrm{dc}^\mathrm{set}-v_\mathrm{dc}$ is used to generate the converter current reference $i_\mathrm{f}^\star$ passed on to the current controller, as illustrated in Fig.~\ref{fig:Hinf_control}. In the proposed approach, however, the conventional static droop relation is replaced by a dynamic controller $R_{\mathcal{H}_\infty}$, which is synthesized to emulate the desired converter behavior introduced in~\cite{Liu2020, Krajacic2026a}. Specifically, $R_{\mathcal{H}_\infty}$ is designed such that the resulting closed-loop output impedance index and current-forming index follow prescribed frequency-domain profiles, which together define the desired grid-forming voltage source behavior, as discussed in the following sections. 

The reference $i_\mathrm{f}^\star$ is subsequently tracked by the inner current controller, producing the voltage reference $v_\mathrm{sw}^\star$ for the Pulse-Width Modulator (PWM), and consequently the switching signal $s_\mathrm{sw}$ for the converter half-bridge. The current controller is a PI controller tuned according to internal model control~\cite{Harnefors1998}:
\begin{equation}\label{eq:cc_bwdth}
    R_\mathrm{i}(s)
    =
    k_\mathrm{pi}\frac{1 + sT_\mathrm{ii}}{sT_\mathrm{ii}},\quad \,
    k_\mathrm{pi} = \omega_{\beta_\mathrm{i}} L_\mathrm{f},\quad \,
    T_\mathrm{ii} = \frac{C_\mathrm{dc} L_\mathrm{f}}{T_\mathrm{d}},
\end{equation}
where $s=j\omega$ denotes the complex frequency-domain operator, and $L_\mathrm{f}$ and $C_\mathrm{dc}$ represent the output filter inductance and capacitance, respectively, as shown in Fig.~\ref{fig:Hinf_control}. Moreover, $\omega_{\beta_\mathrm{i}}$ denotes the current-loop bandwidth and $T_\mathrm{ii}$ is the integral time constant. The bandwidth needs to be set sufficiently below the switching frequency $f_\mathrm{s}$, for example, at $7\%$ of the switching frequency, that is, $\omega_{\beta_\mathrm{i}} = 0.07 \cdot 2\pi f_\mathrm{s}$. Additionally, for the small-signal analysis that follows, the PWM is represented as a mapping from $v_\mathrm{sw}^\star$ to the duty cycle $d$, modeled by a gain of $1/V_\mathrm{dc}$ and a second-order Padé delay $G_\mathrm{t}$ with delay $T_\mathrm{d}=1.25/f_\mathrm{s}$, accounting for sampling, computation, and modulation effects.

\subsection{Source Converter Small-Signal Representation} \label{sec:source_converter_model}

The proposed $\mathcal{H}_\infty$ control design for the I--V type DC-bus control scheme employs the small-signal converter model derived in our previous work~\cite{Krajacic2026a} as the basis for synthesis:
\begin{align}
    \Delta i_{\mathrm{f}} &= G_\mathrm{di}(s) \Delta d + G_\mathrm{oi}(s) \Delta i_\mathrm{o} , \label{eq:ilf_general} \\
    \Delta v_\mathrm{dc} &= G_\mathrm{dv}(s) \Delta d - Z_\mathrm{o}(s) \Delta i_\mathrm{o} , \label{eq:vdc_general}
\end{align}
where the operator $\Delta(\cdot)$ denotes small-signal perturbations around steady-state values, and $G_\mathrm{di}(s)$ and $G_\mathrm{oi}(s)$ represent the transfer functions from $\Delta d$ and $\Delta i_\mathrm{o}$ to $\Delta i_{\mathrm{f}}$, respectively. Moreover, $G_\mathrm{dv}(s)$ denotes the transfer function from $\Delta d$ to $\Delta v_\mathrm{dc}$ and $Z_\mathrm{o}(s)$ corresponds to the open-loop output impedance of the converter. For the specific case of the bidirectional boost converter analyzed in this work and depicted in Fig.~\ref{fig:Hinf_control}, the transfer functions, derived from the averaged model with inductor and switching losses neglected, can be written as:
\begin{subequations}
\begin{align}
    G_\mathrm{di} &= -\frac{I_\mathrm{o}}{D^2}\cdot\frac{1+\frac{Y_\mathrm{dc}V_\mathrm{dc}}{I_\mathrm{o}}}{1+\frac{L_\mathrm{f}Y_\mathrm{dc}}{D^2}s},
    &G_\mathrm{oi} &= \frac{1}{D}\cdot\frac{1}{1+\frac{L_\mathrm{f}Y_\mathrm{dc}}{D^2}s},\\
    G_\mathrm{dv} &= -\frac{V_\mathrm{in}}{D^2}\cdot\frac{1-\frac{L_\mathrm{f}I_\mathrm{f}}{V_\mathrm{in}}s}{1+\frac{L_\mathrm{f}Y_\mathrm{dc}}{D^2}s},
    &Z_\mathrm{o} &= \frac{1}{D^2}\cdot\frac{L_\mathrm{f}s}{1+\frac{L_\mathrm{f}Y_\mathrm{dc}}{D^2}s},
\end{align}
\end{subequations}
where the explicit dependence on $(s)$ is omitted from this point onward for brevity. Here, $Y_\mathrm{dc}=sC_\mathrm{dc}/(1+sC_\mathrm{dc}r_\mathrm{dc})$ denotes the equivalent admittance of the output capacitance, modeled as a capacitor $C_\mathrm{dc}$ in series with its internal resistance $r_\mathrm{dc}$. Furthermore, $I_\mathrm{o}$, $V_\mathrm{in}$, $V_\mathrm{dc}$, and $D$ denote the steady-state output current, input voltage, output voltage, and duty cycle at the operating point, respectively. The duty cycle is defined as the relative on-time of the upper switch with respect to the switching period, while the inductor current is given by the relationship $I_\mathrm{f}=I_\mathrm{o}/D$.

\begin{table*}[t!]
    \centering
    \caption{Device-level performance indices and classification~\cite{Krajacic2026a}.}
    \label{tab:indices}
    \renewcommand{\arraystretch}{1.0}
    \begin{threeparttable}
    \begin{tabularx}{0.99\textwidth}{
        @{}
        >{\centering\arraybackslash}p{0.16\textwidth} 
        >{\centering\arraybackslash}p{0.14\textwidth} 
        >{\centering\arraybackslash}p{0.22\textwidth} 
        >{\centering\arraybackslash}X 
        @{}
    }
        \toprule
        Index & Transfer Function & Definition & Classification \\
        \midrule
        Output Impedance Index (OII)
        &
        $-K_\mathrm{d}\Delta i_\mathrm{o}\mapsto \Delta v_\mathrm{dc}$
        &
        $\displaystyle \mathrm{OII}(\omega)= \frac{Z_\mathrm{out}(j\omega)}{K_\mathrm{d}}$
        &
        $\displaystyle \mathrm{OII}(\omega)= 
            \begin{cases}
                |\mathrm{OII}(\omega)|\leq 1, & \mathrm{grid\!-\!forming}, \\
                |\mathrm{OII}(\omega)|>1, & \mathrm{disturbance\!-\!amplifying}.
            \end{cases}$
        \\[5mm]
        Current-Forming Index (CFI)
        &
        $\Delta i_\mathrm{o}\mapsto \Delta i_\mathrm{c}$
        &
        $\displaystyle \mathrm{CFI}(\omega) = 1-Y_\mathrm{dc}(j\omega)Z_\mathrm{out}(j\omega)$
        &
        $\displaystyle \mathrm{CFI}(\omega)= 
            \begin{cases}
                \mathrm{CFI}(\omega)=1, & \mathrm{current\!-\!following},  \\
                |\mathrm{CFI}(\omega)|< 1, & \mathrm{current\!-\!forming},  \\
                |\mathrm{CFI}(\omega)|\geq1^*, & \mathrm{disturbance\!-\!amplifying}.  
            \end{cases}$
        \\
        \bottomrule
    \end{tabularx}
    \begin{tablenotes}
    \item[*] Includes $|\mathrm{CFI}(\omega)| = 1$ only when $\angle \mathrm{CFI}(\omega) \neq 0$.
    \end{tablenotes}
\end{threeparttable}
\end{table*}

\begin{figure}[b!]
    \centering
    \includegraphics[]{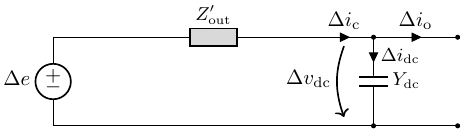}
    \caption{Source converter small-signal equivalent circuit used for output voltage regulation analysis.}
    \label{fig:equivalent_circuit}
\end{figure}

To analyze the output voltage regulation of a source converter, the small-signal equivalent circuit must explicitly capture the capacitor and control-related dynamics, since this regulation is closely linked to the current flowing into the output capacitance~\cite{Deng2024}. With this in mind, the voltage control dynamics of the I--V $\mathcal{H}_\infty$ source converter depicted in Fig.~\ref{fig:Hinf_control} are analyzed using the small-signal circuit shown in Fig.~\ref{fig:equivalent_circuit}. Here, the converter is modeled as a non-ideal voltage source with a series output impedance $Z_\mathrm{out}'$ and a parallel capacitive admittance $Y_\mathrm{dc}$. Together, these two form the conventional output impedance $Z_\mathrm{out}$ of the source converter, which encompasses both the converter topology and the analyzed I--V $\mathcal{H}_\infty$ control scheme. 

The analytical expression for $Z_\mathrm{out}$ is derived following the approach detailed in~\cite{Krajacic2026a, Stanojev2026}. In brief, the small-signal open-loop converter model~\eqref{eq:ilf_general}--\eqref{eq:vdc_general} is extended to incorporate the I--V~$\mathcal{H}_\infty$ control algorithm by expressing the duty-cycle deviations $\Delta d$ as a linear combination of $\Delta v_\mathrm{dc}$ and $\Delta i_\mathrm{f}$. These are the only measured feedback signals of the control scheme, serving as inputs to the $\mathcal{H}_\infty$ and current controllers, respectively, while all references are constant and thus vanish in the small-signal model. This yields the closed-loop transfer function from $-\Delta i_\mathrm{o}$ to $\Delta v_\mathrm{dc}$, i.e., $Z_\mathrm{out}$, whose explicit expression is given subsequently in~\eqref{eq:Zout_Hinf}. Then, using the convention from Fig.~\ref{fig:equivalent_circuit}, the series output impedance capturing the converter and control dynamics follows as $Z_\mathrm{out}'=\left(1/Z_\mathrm{out}-Y_\mathrm{dc}\right)^{-1}$. The corresponding circuit equations can be written as:
\begin{align}
    \Delta i_{\mathrm{dc}} &=Y_\mathrm{dc}\Delta v_\mathrm{dc} =  \Delta i_\mathrm{c} - \Delta i_\mathrm{o} , \label{eq:Cdc} \\
    \Delta v_\mathrm{dc} &= \Delta e - Z_\mathrm{out}'\Delta i_\mathrm{c} , \label{eq:vdc}
\end{align}
where $\Delta i_\mathrm{c}$ represents the converter current deviations, $\Delta i_\mathrm{o}$ the output current variations, and $\Delta i_{\mathrm{dc}}$ denotes the current perturbations of the output capacitor. Thus, in this model, the behavior of the output voltage fluctuation $\Delta v_\mathrm{dc}$ is directly influenced by $\Delta i_\mathrm{dc}$ and the characteristics of the impedance $Z_\mathrm{out}'$ and admittance $Y_\mathrm{dc}$.

\subsection{Characterization of Device-Level Behavior} \label{sec:GFM_characterization}
The device-level performance indices introduced in our previous work~\cite{Krajacic2026a}, derived from~\eqref{eq:Cdc}--\eqref{eq:vdc} with $\Delta e=0$, are summarized in Table~\ref{tab:indices}. These indices quantify the converter's ability to regulate its output voltage $v_\mathrm{dc}$ and internal current $i_\mathrm{c}$ in the presence of output current $i_\mathrm{o}$ deviations. Each index characterizes a different aspect of the converter response to disturbances, providing a comprehensive assessment of device-level voltage-control performance.

The OII quantifies the converter's ability to regulate its terminal voltage in response to output-current disturbances. Since the droop gain $K_\mathrm{d}$ represents the ratio between the maximum permissible voltage deviation and output-current deviation, $|\mathrm{OII}(\omega)|\leq1$ indicates that the converter limits the resulting voltage deviation within the prescribed bound. Conversely, $|\mathrm{OII}(\omega)|>1$ corresponds to disturbance-amplifying behavior. In addition, provided the standard strict-passivity conditions are satisfied~\cite{Khalil2002}, the OII phase can be used to assess passivity. Specifically, passivity is ensured if the OII phase remains within $\pm 90^\circ$ over all frequencies, thus satisfying the passivity-based stability condition~\cite{Lazarevic2025}.

The CFI characterizes how the internal current of the converter responds to output current disturbances relative to the contribution of the output capacitance. For $\mathrm{CFI}(\omega)=1$, the converter current follows the output-current variation, resulting in current-following behavior and no capacitor-current deviation, i.e., $\Delta i_\mathrm{dc}=0$. In contrast, $|\mathrm{CFI}(\omega)|<1$ indicates current-forming behavior, where a larger share of the disturbance is contained by the output capacitance, while values outside these regions result in disturbance-amplifying~behavior.

\subsection{Desired Grid-Forming Behavior} \label{sec:desired}
Building on the indices summarized in Table~\ref{tab:indices}, we leverage their characteristics to formulate conditions for desired grid-forming behavior introduced in~\cite{Krajacic2026a}. As discussed in Section~\ref{sec:source_converter_model}, the converter output impedance $Z_\mathrm{out}$ is given by the parallel combination of the output capacitor admittance $Y_\mathrm{dc}$ and the impedance $Z_\mathrm{out}'$ associated with the converter control dynamics. Ideally, $Z_\mathrm{out}'$ exhibits resistive droop behavior, i.e., $Z_\mathrm{out}'=K_\mathrm{d}$, while the output capacitance attenuates disturbances at higher frequencies, resulting in $Z_\mathrm{out}=K_\mathrm{d}||(1/Y_\mathrm{dc})$~\cite{Liu2020}. Accordingly, the desired converter output impedance is defined as:
    \begin{equation}
    Z_\mathrm{out,d}(j\omega)
    =
    \frac{K_\mathrm{d}}
    {1+Y_\mathrm{dc}(j\omega)K_\mathrm{d}},
    \label{eq:Z_desired}
\end{equation}
where the corresponding crossover frequency $\omega_\mathrm{c}$ between resistive and capacitive $Z_\mathrm{out}$ can be approximated for $K_\mathrm{d}\gg r_\mathrm{dc}$:
\begin{equation}
    \omega_\mathrm{c}
    =
    \frac{1}{C_\mathrm{dc}K_\mathrm{d}}.
    \label{eq:fbd}
\end{equation}

Substituting~\eqref{eq:Z_desired} into the OII and CFI definitions from Table~\ref{tab:indices} yields the equivalent desired frequency response for both indices, namely:
\begin{equation}
    \mathrm{OII}_\mathrm{d}(\omega)
    =
    \mathrm{CFI}_\mathrm{d}(\omega)
    =
    \frac{1}
    {1+Y_\mathrm{dc}(j\omega)K_\mathrm{d}}.
\label{eq:OII_CFI_d}
\end{equation}
Hence, the desired behavior transitions from current-following operation at low frequencies, where $\mathrm{CFI}_\mathrm{d}\approx1$, toward current-forming behavior at higher frequencies as the output capacitance becomes dominant and therefore $|\mathrm{CFI}_\mathrm{d}(\omega)|<1$. However, because $|\mathrm{OII}_\mathrm{d}(\omega)| = |\mathrm{CFI}_\mathrm{d}(\omega)|\leq 1$ the behavior stays grid-forming over the entire frequency range. 

\section{\texorpdfstring{$\mathcal{H}_\infty$}{H-infinity} Control Design} \label{sec:Hinf_control}

The proposed I--V $\mathcal{H}_\infty$ control design shown in Fig.~\ref{fig:Hinf_control} is represented by the small-signal control diagram in Fig.~\ref{fig:Hinf_Diagram}. Here, the converter model~\eqref{eq:ilf_general}--\eqref{eq:vdc_general} is depicted with grey blocks, while the inner current controller $R_\mathrm{i}$ and PWM are highlighted in green. The outer $\mathcal{H}_\infty$ controller $R_{\mathcal{H}_\infty}$, which forms the core of the $\mathcal{H}_\infty$ control design presented in this section, is highlighted in blue. Finally, the targeted device-level performance indices and the corresponding weighting functions are depicted in red.

\subsection{Tracking of Desired Grid-Forming Behavior}
\begin{figure}[t!]
    \centering
    \includegraphics[width=\linewidth]{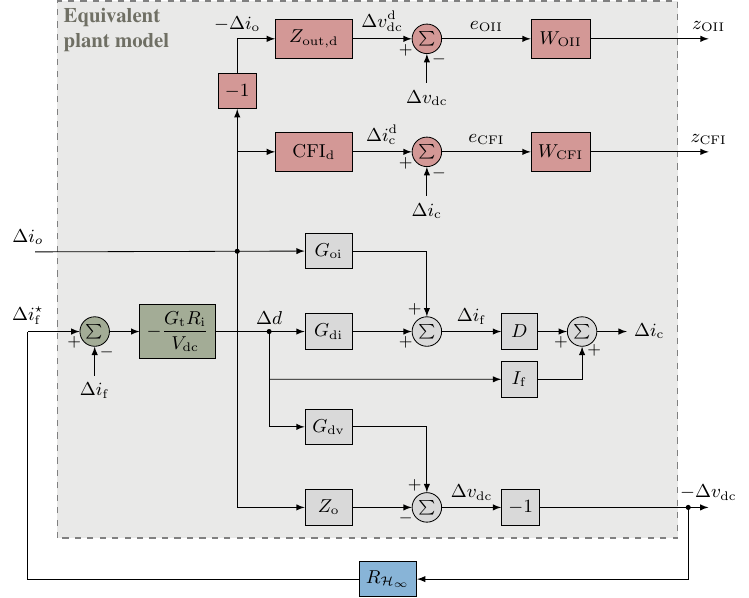}
    \caption{Small-signal control diagram of I--V $\mathcal{H}_\infty$ control from Fig.~\ref{fig:Hinf_control} used for optimization with current controller and PWM (green), converter model (grey), error blocks (red) and $R_{\mathcal{H}_\infty}$ controller (blue).}
    \label{fig:Hinf_Diagram}
\end{figure}

The synthesis objective is to achieve the desired OII and CFI characteristics~\eqref{eq:OII_CFI_d} by tuning the $R_{\mathcal{H}_\infty}$ controller transfer function. Since $\mathrm{OII}(\omega)=Z_\mathrm{out}/K_\mathrm{d}$ as per Table~\ref{tab:indices} and assuming the droop gain is fixed, tracking the desired OII characteristic is equivalent to tracking the desired output impedance. Therefore, tracking errors are defined as deviations of the achieved output impedance and CFI profiles from their desired values. These can be expressed in terms of terminal signal deviations for the disturbance $\Delta i_\mathrm{o}$, shown by red blocks in Fig.~\ref{fig:Hinf_Diagram}:
\begin{alignat}{2}
    e_\mathrm{OII}
    &= \Delta v_\mathrm{dc}^\mathrm{d} - \Delta v_\mathrm{dc}
    &{}={}& \left(Z_\mathrm{out,d} - Z_\mathrm{out}\right)\cdot\bigl(-\Delta i_\mathrm{o}\bigr),
    \label{eq:eZ} \\
    e_\mathrm{CFI}
    &= \Delta i_\mathrm{c}^\mathrm{d} - \Delta i_\mathrm{c}
    &{}={}& \left(\mathrm{CFI}_\mathrm{d} - \mathrm{CFI}\right)\,\Delta i_\mathrm{o},
    \label{eq:eCFI}
\end{alignat}
where $\Delta v_\mathrm{dc}^\mathrm{d} = Z_\mathrm{out,d}\cdot(-\Delta i_\mathrm{o})$ represents the desired voltage response of the converter, $\Delta i_\mathrm{c}^\mathrm{d} = \mathrm{CFI}_\mathrm{d}\,\Delta i_\mathrm{o}$ the desired converter current deviations, and $Z_\mathrm{out,d}$ and $\mathrm{CFI}_\mathrm{d}$ are given by~\eqref{eq:Z_desired} and~\eqref{eq:OII_CFI_d}, respectively. 

\subsection{Performance Output Weighting Functions} \label{sec:Hinf_W}

The performance output vector $\mathbf{z}$ is constructed by passing the tracking errors~\eqref{eq:eZ}--\eqref{eq:eCFI} through the frequency-selective weighting functions $W_\mathrm{OII}$ and $W_\mathrm{CFI}$, respectively:
\begin{equation}
    \mathbf{z} =
    \begin{bmatrix}
        z_\mathrm{OII} \\
        z_\mathrm{CFI}
    \end{bmatrix}
    =
    \begin{bmatrix}
        W_\mathrm{OII} \, e_\mathrm{OII}\\
        W_\mathrm{CFI} \, e_\mathrm{CFI}
    \end{bmatrix}.
    \label{eq:weighted_errors}
\end{equation}
Both weights are realized as first-order shaping filters:
\begin{equation}
    W(s) = W_\infty \cdot
    \frac{s + \omega_\mathrm{a} / W_\infty}
         {s + \omega_\mathrm{a} / W_\mathrm{o}},
    \label{eq:weight_form}
\end{equation}
with DC gain $W_\mathrm{o}$, high-frequency gain $W_\infty$, and asymptotic $0\,\mathrm{dB}$ crossover $\omega_\mathrm{a}$, where $W_\mathrm{o}>1>W_\infty$. The same form is used for both weights, with superscripts denoting the respective weight. In this formulation, the DC gain $W_\mathrm{o}\gg1$ strongly penalizes tracking errors at low frequencies, thereby enforcing accurate steady-state tracking of the desired converter response. Conversely, $W_\infty<1$ relaxes the penalty at high frequencies, which lie beyond the current-control bandwidth and where the response is increasingly governed by the output capacitance, thus reducing the incentive for the synthesis to aggressively track the desired profile in a region where the controller has limited authority. In this regard, $W_\infty^\mathrm{CFI}$ has little influence on the resulting controller because the output capacitance predominantly determines the current-forming behavior in this frequency range~\cite{Krajacic2026a}. In contrast, $W_\infty^\mathrm{OII}$ balances the OII magnitude overshoot against low-to-mid-frequency tracking and is therefore set slightly below unity (see Table~\ref{tab:parameters}).

The asymptotic crossover is calculated as
\begin{equation}
    \omega_\mathrm{a} = \omega_\mathrm{o}\sqrt{\frac{1-W_\infty^{2}}{1-1/W_\mathrm{o}^{2}}},
    \label{eq:W_crossover}
\end{equation}
such that each weight crosses $0\,\mathrm{dB}$ exactly at its own crossover frequency $\omega_\mathrm{o}=2\pi f_\mathrm{o}$. Since the two weights are designed independently, two crossover frequencies result, namely $f_\mathrm{o}=f_\mathrm{OII}$ for $W_\mathrm{OII}$ and $f_\mathrm{o}=f_\mathrm{CFI}$ for $W_\mathrm{CFI}$, which are referenced to characteristic converter frequencies according to:
\begin{equation}
    f_\mathrm{OII} = \alpha_\mathrm{OII} f_{\beta_\mathrm{i}},
    \qquad
    f_\mathrm{CFI} = \alpha_\mathrm{CFI} f_\mathrm{c},
    \label{eq:W_bandwidth}
\end{equation}
where $\alpha_\mathrm{OII}$ and $\alpha_\mathrm{CFI}$ are tunable scaling factors. Here, $f_{\beta_\mathrm{i}}$ denotes the current-control bandwidth~\eqref{eq:cc_bwdth}, while $f_\mathrm{c}$, defined in~\eqref{eq:fbd}, corresponds to the transition of the desired output impedance from resistive to capacitive behavior and, consequently, from current-following to current-forming operation. The influence and selection of the scaling factors are further elaborated in Section~\ref{sec:oii_analysis}.

\subsection{Methodology} \label{sec:Hinf_plant_s}

\begin{figure}[b!]
    \centering
    \includegraphics[width=\linewidth]{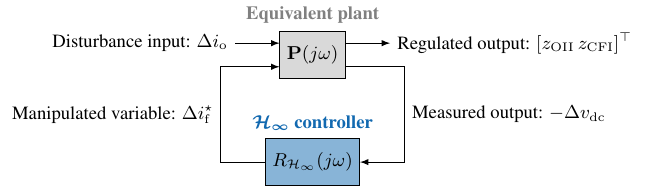}
    \caption{Partitioned-form representation of plant model from Fig.~\ref{fig:Hinf_Diagram}.}
    \label{fig:Hinf_Diagram_Simplified}
\end{figure}

The control diagram shown in Fig.~\ref{fig:Hinf_Diagram} can be represented by its partitioned form, depicted in Fig.~\ref{fig:Hinf_Diagram_Simplified}. In this representation, the generalized plant $\mathbf{P}(j\omega)$ maps the disturbance and control inputs $(\Delta i_\mathrm{o}, \Delta i_\mathrm{f}^\star)$ to the performance and measurement outputs $(\mathbf{z},-\Delta v_\mathrm{dc})$, i.e.,
\begin{equation}
    \resizebox{0.89\linewidth}{!}{$
    \displaystyle
    \begin{bmatrix}
        \mathbf{z} \\
        -\Delta v_\mathrm{dc}
    \end{bmatrix}
    =
    \mathbf{P}(j\omega)
    \begin{bmatrix}
        \Delta i_\mathrm{o} \\
        \Delta i_\mathrm{f}^\star
    \end{bmatrix}
    =
    \begin{bmatrix}
        \mathbf{P}_{11}(j\omega) & \mathbf{P}_{12}(j\omega) \\
        \mathbf{P}_{21}(j\omega) & \mathbf{P}_{22}(j\omega)
    \end{bmatrix}
    \begin{bmatrix}
        \Delta i_\mathrm{o} \\
        \Delta i_\mathrm{f}^\star
    \end{bmatrix}.
    $}
    \label{eq:Hinf_generalized_plant}
\end{equation}
The control law is given by
\begin{equation}
    \Delta i_\mathrm{f}^\star = R_{\mathcal{H}_\infty}(j\omega)\cdot\bigl(-\Delta v_\mathrm{dc}\bigr), \label{eq:Hinf_controller}
\end{equation}
where $R_{\mathcal{H}_\infty}(j\omega)$ denotes the $\mathcal{H}_\infty$ controller transfer function to be designed. The negative sign stems from the controller input being the output-voltage deviation $v_\mathrm{dc}^\mathrm{set}-v_\mathrm{dc}$, whose small-signal counterpart is $-\Delta v_\mathrm{dc}$. This is accounted for by the gain of $-1$ in the equivalent plant in Fig.~\ref{fig:Hinf_Diagram}.

By substituting the control law~\eqref{eq:Hinf_controller} into the generalized plant representation~\eqref{eq:Hinf_generalized_plant}, the dependence of the regulated output $\mathbf{z}$ on the disturbance input $\Delta i_\mathrm{o}$ can be expressed as
\begin{equation}
    \mathbf{z} = \mathbf{F}_\ell(\mathbf{P},R_{\mathcal{H}_\infty})\,\Delta i_\mathrm{o},
    \label{eq:Hinf_LFT_mapping}
\end{equation}
where $\mathbf{F}_\ell(\mathbf{P},R_{\mathcal{H}_\infty})$ denotes the lower linear fractional transformation of the generalized plant $\mathbf{P}$~\cite{Zhou1996}:
\begin{equation}
    \resizebox{0.89\linewidth}{!}{$
    \displaystyle
    \mathbf{F}_\ell(\mathbf{P},R_{\mathcal{H}_\infty})
    =
    \mathbf{P}_{11}
    +
    \mathbf{P}_{12}\,R_{\mathcal{H}_\infty}
    \left(
        \mathbf{I}
        -
        \mathbf{P}_{22}\,R_{\mathcal{H}_\infty}
    \right)^{-1}
    \mathbf{P}_{21}.
    $}
    \label{eq:Hinf_LFT_definition}
\end{equation}
Here, $\mathbf{P}_{ij}$ denotes the corresponding sub-blocks of the generalized plant $\mathbf{P}$ defined in~\eqref{eq:Hinf_generalized_plant}, while the explicit dependence on $(j\omega)$ is omitted for brevity.

The objective of the $\mathcal{H}_\infty$ control design is to determine a stabilizing controller $R_{\mathcal{H}_\infty}$ such that the closed-loop transfer function from the disturbance input $\Delta i_\mathrm{o}$ to the regulated output $\mathbf{z}$ is minimized in the $\mathcal{H}_\infty$ sense. This objective can be expressed as:
\begin{equation}
    \resizebox{0.89\linewidth}{!}{$
    \displaystyle
    \min_{R_{\mathcal{H}_\infty}}
    \left\| \mathbf{F}_\ell(\mathbf{P},R_{\mathcal{H}_\infty}) \right\|_\infty
    =
    \min_{R_{\mathcal{H}_\infty}}
    \sup_{\omega}
    \bar{\sigma}
    \!\left(
    \mathbf{F}_\ell(\mathbf{P},R_{\mathcal{H}_\infty})(j\omega)
    \right),
    $}
\end{equation}
where $\bar{\sigma}(\cdot)$ denotes the maximum singular value.
Solving this optimization problem yields the $R_{\mathcal{H}_\infty}$ transfer function that minimizes the worst-case amplification of the weighted error signals, thus enforcing the desired OII and CFI characteristics over the frequency range of interest.

\begin{table*}[b!]
    \centering
    \small
    \begin{equation}
        Z_\mathrm{out}
        (j\omega)=
        \dfrac{s L_\mathrm{f} + G_\mathrm{t} R_\mathrm{i}}
        {\left( Y_\mathrm{dc} + \dfrac{D^2}{s L_\mathrm{f}} \right)
         \left( s L_\mathrm{f} + G_\mathrm{t} R_\mathrm{i} \right) 
        + \dfrac{G_\mathrm{t} R_\mathrm{i}}{V_\mathrm{dc}}
        \left( I_\mathrm{f} - \dfrac{V_\mathrm{dc}D}{s L_\mathrm{f}} \right)
        \left( D - R_{\mathcal{H}_\infty} s L_\mathrm{f} \right)
    }
    \label{eq:Zout_Hinf}
    \tag{25}
    \end{equation}
\end{table*}

\subsection{Controller Constraints} \label{sec:Hinf_design}

We constrain the proposed $R_{\mathcal{H}_\infty}$ controller to a fixed-order rational transfer function of prescribed numerator degree $n_z$ and denominator degree $n_p$:
\begin{equation}
    R_{\mathcal{H}_\infty}
    \;=\;
    \frac{b_{n_z} s^{n_z} + \cdots + b_0}
         {a_{n_p} s^{n_p} + \cdots + a_0}
    \;\in\;
    \mathcal{K}_{n_z,n_p},
    \label{eq:Hinf_structure}
\end{equation}
where $n_z\leq n_p$ ensures a proper, and hence realizable, controller. Beyond this condition, the orders are free design choices. In practice, however, the transfer function order may be limited by implementation, since higher orders increase the computational burden on the control platform.

In addition to the impedance shaping objective, a passivity requirement is imposed on the resulting OII phase, which constitutes a sufficient condition for stable DC microgrid interconnection as described in Section~\ref{sec:GFM_characterization}. This condition is expressed as:
\begin{equation}
    -\pi/2  - \varepsilon_\mathrm{p} \leq 
    \angle{\mathrm{OII}(\omega)} 
    \leq \pi/2  +\varepsilon_\mathrm{p},
    \quad \forall\,\omega \in [0,\infty),
    \label{eq:passivity_constraint}
\end{equation}
where $\varepsilon_\mathrm{p}$ is a passivity tolerance. The default choice of $\varepsilon_\mathrm{p} = 0$ enforces passivity (OII phase within $\pm 90^\circ$ at all frequencies), while $\varepsilon_\mathrm{p} < 0$ introduces an additional phase margin of $|\varepsilon_\mathrm{p}|$ to the passivity boundary. Conversely, $\varepsilon_\mathrm{p} > 0$ relaxes the constraint and accommodates plant models for which enforcing $\varepsilon_\mathrm{p}$ equal to zero may yield an infeasible optimization problem.

\subsection{Controller Synthesis} 
The complete synthesis problem is stated as
\begin{align}
\min_{R_{\mathcal{H}_\infty}\,\in\,\mathcal{K}_{n_z,n_p}}
    &\quad
    \bigl\| \mathbf{F}_\ell\!\left(\mathbf{P},\,R_{\mathcal{H}_\infty}\right) \bigr\|_\infty \label{eq:opt_obj} \\[3pt]
    \text{subject to}
    &\quad
    -\pi/2  - \varepsilon_\mathrm{p} \leq 
    \angle{\mathrm{OII}(\omega)} 
    \leq \pi/2  +\varepsilon_\mathrm{p}, \label{eq:opt_constraint}
\end{align}
for $\omega \in [0,\infty)$, where $\mathbf{F}_\ell(\mathbf{P}, R_{\mathcal{H}_\infty})$ from~\eqref{eq:opt_obj} is the lower linear fractional transformation defined by~\eqref{eq:Hinf_LFT_definition}, while~\eqref{eq:opt_constraint} is the passivity constraint detailed in the previous section. The $\mathcal{H}_\infty$ norm is computed using the algorithm from~\cite{Bruinsma1990} and is non-differentiable at frequencies where its maximum singular value is attained. The problem is solved via non-smooth optimization using Clarke subgradients~\cite{Apkarian2006} and is performed entirely offline. Hardware implementation requires only evaluation of the synthesized first-order transfer function. The constraint~\eqref{eq:opt_constraint} is enforced as a hard constraint in the synthesis, so that the obtained controller satisfies it a priori rather than having to be verified a posteriori~\cite{Apkarian2014}.

\section{Frequency-Domain Analysis} \label{sec:results_freq}

\subsection{Output Impedance Analysis and Tuning Guidelines} \label{sec:oii_analysis}

\begin{figure}[t!]
	\centering
	\includegraphics[width=\linewidth]{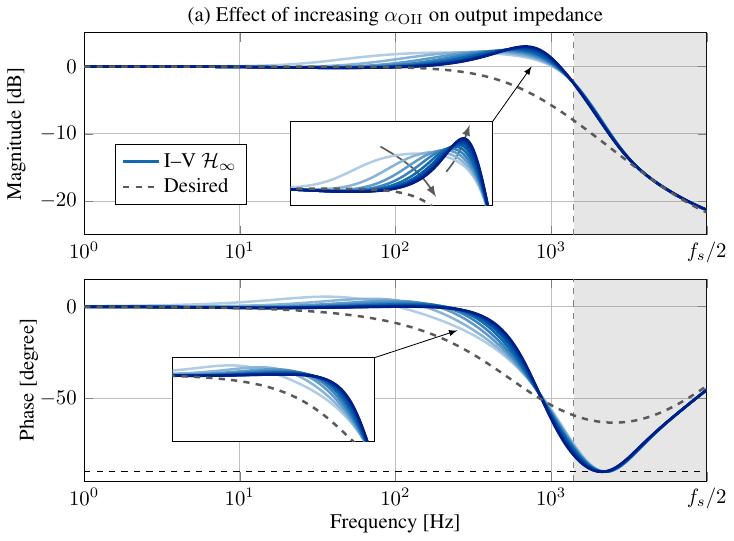}
	\includegraphics[width=\linewidth]{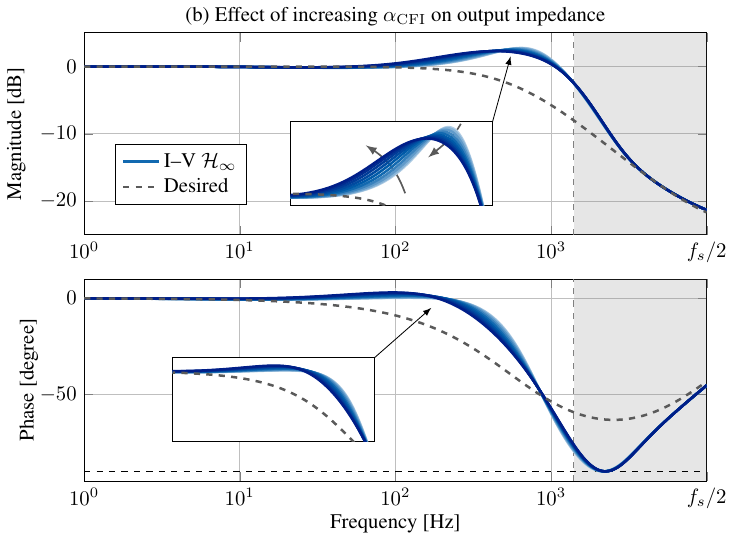}
    \caption{Bode diagram of the output impedance for increasing values of (a) $\alpha_\mathrm{OII}$ and (b) $\alpha_\mathrm{CFI}$, where darker shades of blue and the directions of the grey arrows in the zoomed-in insets correspond to larger parameter values. Hereafter, the grey shaded area refers to the frequency range above the current-control bandwidth $f_{\beta_\mathrm{i}}$, with the frequency range shown extending up to half the converter switching frequency $f_\mathrm{s}/2$.}
	\label{fig:Zout_Hinf_sweep}
\end{figure}

Based on the control diagram in Fig.~\ref{fig:Hinf_Diagram}, the analytical expression for the output impedance of the I--V~$\mathcal{H}_\infty$ control algorithm can be derived and is given by~\eqref{eq:Zout_Hinf}. With this expression at hand, we first examine how the weighting functions, in particular the scaling factors $\alpha_\mathrm{OII}$ and $\alpha_\mathrm{CFI}$, shape the closed-loop impedance. Each factor is swept from 0.2 to 2 while the other is held at unity. The detailed system parameters and operating points used for the control design are provided in Table~\ref{tab:parameters}, and the $\mathcal{H}_\infty$ synthesis is carried out in MATLAB using the Robust Control Toolbox \cite{MATLAB}. The resulting output impedances are shown in Fig.~\ref{fig:Zout_Hinf_sweep}.

\begin{table}[t!]
	\centering
	\caption{Parameters of the $\mathcal{H}_\infty$ synthesis and experimental setup.}
	\label{tab:parameters}
    \begin{threeparttable}
    \renewcommand{\arraystretch}{1.10}
	\begin{tabular}{lcr}
		\toprule
		Quantity & Symbol & Value \\
		\midrule
        DC-bus voltage setpoint             & $v_\mathrm{dc}^\mathrm{set}$ & \SI{350}{\volt}          \\
		Boost input voltage                 & $V_\mathrm{in}$              & \SI{200}{\volt}          \\
        Boost initial active power output         & $p_\mathrm{o}$  & \SI{1}{\kilo\watt}       \\
        Boost filter inductance             & $L_\mathrm{f}$               & \SI{2.2}{\milli\henry}   \\
        Boost output capacitance                 & $C_\mathrm{dc}$              & \SI{250}{\micro\farad}   \\
        Boost output capacitor series resistance*  & $r_\mathrm{dc}$              & \SI{60}{\milli\ohm}      \\
        \midrule
        PC Buck output-power setpoint        & $p_\mathrm{o}^\mathrm{set}$  & \SI{1}{\kilo\watt}          \\
        PC Buck filter inductance            & $L_\mathrm{f}$             & \SI{2.2}{\milli\henry}     \\
        PC Buck output voltage               & $V_\mathrm{out}$              & \SI{200}{\volt}     \\
        \midrule
        VC Buck output-voltage setpoint        & $v_\mathrm{o}^\mathrm{set}$  & \SI{100}{\volt}          \\
        VC Buck filter inductance              & $L_\mathrm{f}$             & \SI{1}{\milli\henry}     \\
        VC Buck filter capacitance             & $C_\mathrm{f}$             & \SI{100}{\micro\farad}   \\
        VC Buck output load                    & $R_\mathrm{load}$             & \SI{9.5}{\ohm}   \\
        \midrule
        Switching frequency                 & $f_\mathrm{s}$               & \SI{20}{\kilo\hertz}     \\
        Total loop delay                    & $T_\mathrm{d}$               & \SI{62.5}{\micro\second} \\
        \midrule
        Current-loop bandwidth              & $f_{\beta_\mathrm{i}}$       & \SI{1400}{\hertz}        \\
        Desired impedance corner frequency  & $f_\mathrm{c}$          & \SI{637}{\hertz}         \\
        Desired droop gain                  & $K_\mathrm{d}$               & \SI{1}{\ohm} \\
        \midrule
        Weight DC gain                      & $W_\mathrm{o}$                  & $10^4$                   \\
        Weight high-frequency gains         & $W_{\infty}^\mathrm{OII}$/$W_{\infty}^\mathrm{CFI}$ & $0.9$/$0.1$ \\
        Weight scaling factors              & $\alpha_\mathrm{OII}$/$\alpha_\mathrm{CFI}$ & $1$/$1$    \\
        Controller order                    & $n_z$/$n_p$                    & $1$/$1$                    \\
        Passivity tolerance                 & $\varepsilon_\mathrm{p}$     & $0$                      \\
		\bottomrule
	\end{tabular}
    \begin{tablenotes}
    \item[*] Estimated for the purpose of $\mathcal{H}_\infty$ synthesis.
    \end{tablenotes}
    \end{threeparttable}
\end{table}

\begin{figure}[b!]
    \centering
    \includegraphics[width=\linewidth]{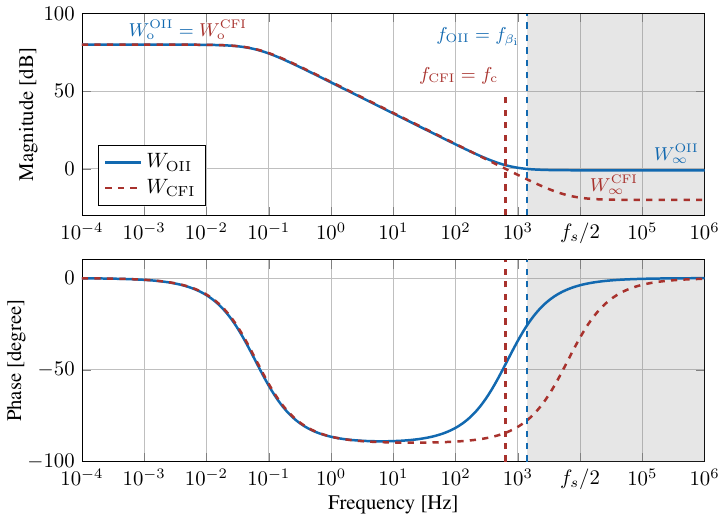}
    \caption{Weighting functions $W_\mathrm{OII}$ and $W_\mathrm{CFI}$ used for the $\mathcal{H}_\infty$ synthesis.}
    \label{fig:Hinf_W}
\end{figure}

\textit{Effect of increasing $\alpha_\mathrm{OII}$:}
Increasing $\alpha_\mathrm{OII}$ shifts the OII weighting crossover $f_\mathrm{OII}$ to higher frequencies, penalizing the impedance-tracking error over a wider band. The shaped impedance consequently follows the desired resistive plateau further into the mid-frequency range. Beyond unity, however, $f_\mathrm{OII}$ exceeds the current-control bandwidth $f_{\beta_\mathrm{i}}$, i.e., the synthesis is driven to act beyond the converter control bandwidth. The additional weighting therefore yields only a marginal improvement in mid-frequency tracking, while the magnitude overshoot in the higher-frequency range increases accordingly. Hence, $\alpha_\mathrm{OII}$ is set to 1, which places $f_\mathrm{OII}$ at $f_{\beta_\mathrm{i}}$ as a reasonable trade-off, therefore emphasizing accurate impedance tracking below the finite current-control bandwidth.

\textit{Effect of increasing $\alpha_\mathrm{CFI}$:}
The scaling factor $\alpha_\mathrm{CFI}$ exhibits the opposite trade-off: raising it extends the band over which the converter must reproduce the desired current-forming behavior. This is obtained at the expense of low-to-mid-frequency accuracy, since above $f_\mathrm{c}$ the output capacitance is intended to supply the transient current component. In other words, forcing the controller to continue tracking the CFI target beyond this frequency yields an excessively current-following response with respect to the CFI. Therefore, $\alpha_\mathrm{CFI}=1$ is retained, setting the crossover to that of the desired impedance, i.e., $f_\mathrm{CFI}=f_\mathrm{c}$ as defined in~\eqref{eq:fbd}. This frequency marks the transition of the output impedance from resistive to capacitive behavior and, consequently, the transition from current-following to current-forming operation. Nevertheless, both scaling factors remain tunable design parameters and can be adjusted to accommodate different converter characteristics, operating conditions, and control objectives.

\begin{figure}[t!]
    \centering
    \includegraphics[width=\linewidth]{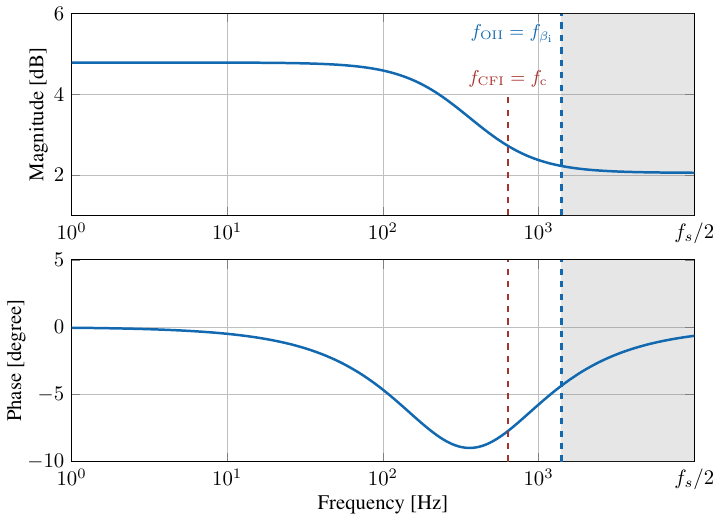}
    \caption{Resulting $R_{\mathcal{H}_\infty}$ transfer function with $n_z=1$ and $n_p=1$.}
    \label{fig:Hinf_K}
\end{figure}

\begin{figure}[b!]
    \centering
    \includegraphics[width=\linewidth]{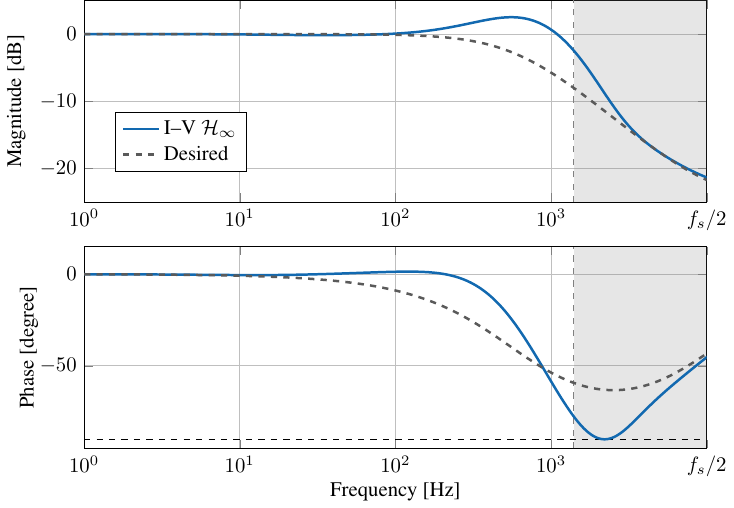}
    \caption{Resulting I--V $\mathcal{H}_\infty$ analytical output impedance \eqref{eq:Zout_Hinf}.}
    \label{fig:Zout_Hinf}
\end{figure}

\textit{Resulting output impedance:}
With the scaling factors fixed at $\alpha_\mathrm{OII}=\alpha_\mathrm{CFI}=1$, the weighting functions take the form shown in Fig.~\ref{fig:Hinf_W}, with DC and high-frequency gains summarized in Table~\ref{tab:parameters}. The resulting $R_{\mathcal{H}_\infty}$ controller obtained from the $\mathcal{H}_\infty$ synthesis for $n_z=n_p=1$ is shown in Fig.~\ref{fig:Hinf_K}. Furthermore, Fig.~\ref{fig:Zout_Hinf} presents the resulting I--V $\mathcal{H}_\infty$ output impedance trajectory for the selected parameters compared against the desired characteristic~\eqref{eq:Z_desired} in dashed grey. As observed, the two curves closely match in the low- and mid-frequency regions, while a slight deviation arises at mid-to-high frequencies due to the finite controller bandwidth, fixed size of the output capacitance, and the control loop delay.

\begin{figure*}[t!]
    \centering
    \includegraphics[width=\textwidth]{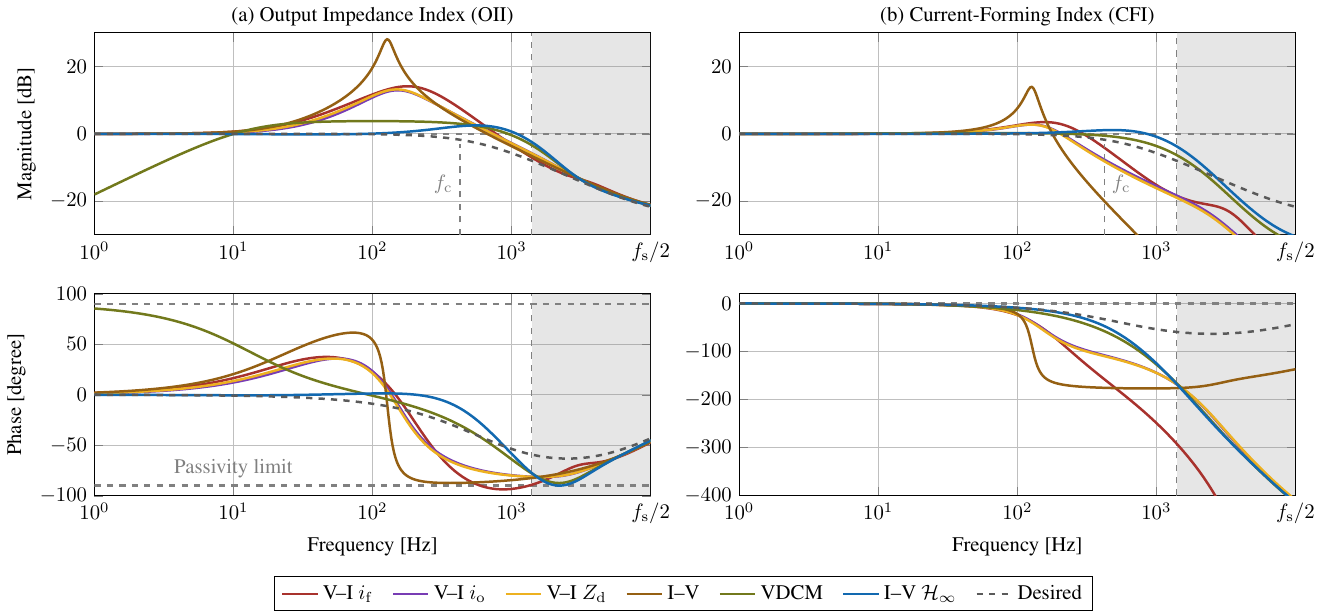}
    \caption{Device-level performance indices for different source converter control algorithms: (a) Output Impedance Index (OII), (b) Current-Forming Index~(CFI).}
    \label{fig:index_plots}
\end{figure*}

\subsection{Performance Comparison} \label{sec:device_analysis}

We benchmark the novel I--V $\mathcal{H}_\infty$ control algorithm against several representative droop-based control strategies from recent literature~\cite{Krajacic2026a}. These include V--I droop with inductor current measurement (V--I~$i_{\mathrm{f}}$)~\cite{Jin2017b}, V--I droop with output current measurement and a constant droop coefficient (V--I~$i_{\mathrm{o}}$)~\cite{Jin2017a}, and frequency-dependent droop control (V--I~$Z_{\mathrm{d}}$)~\cite{Liu2020}. For I--V droop control, the controller proposed in~\cite{Lazarevic2025} is implemented and denoted as I--V. As a representative virtual DC machine control strategy, the power-loop-free variant introduced in~\cite{Na2022} is analyzed and denoted as VDCM. Finally, the desired converter behavior (Desired) is implemented as an ideal voltage source with a resistive--capacitive output impedance, as defined by~\eqref{eq:Z_desired}, and is used as a reference throughout the analysis. All benchmarked controllers are evaluated using the same converter parameters and operating points as in Table~\ref{tab:parameters}, with controller-specific parameters selected according to the tuning procedures of the corresponding references.

The OII trajectories in Fig.~\ref{fig:index_plots}(a) show that the I--V $\mathcal{H}_\infty$ controller matches the desired behavior most closely. Its excursion into the disturbance-amplifying region is confined to a narrow band around the corner frequency $f_\mathrm{c}$ and is limited to $2.5\,\mathrm{dB}$, against $3.8\,\mathrm{dB}$ for the VDCM and between $13$ and $28\,\mathrm{dB}$ for the remaining droop-based variants. The latter thus amplify voltage deviations over a broad frequency range, leading to pronounced voltage-magnitude over- and undershoots under disturbances~\cite{Liu2020}. Moreover, the OII phase of the I--V $\mathcal{H}_\infty$ controller remains within $\pm90^\circ$ across the entire frequency range, so the passivity requirement~\eqref{eq:passivity_constraint} is met by construction, whereas among the benchmarked algorithms the V--I~$i_\mathrm{f}$ variant exhibits a phase excursion beyond $-90^\circ$.

The CFI trajectories in Fig.~\ref{fig:index_plots}(b) confirm the OII trends. All algorithms operate in the current-following region at low frequencies, while their transition toward current-forming behavior differs: the I--V $\mathcal{H}_\infty$ controller follows the desired transition most closely, remaining nearest to $\mathrm{CFI}_\mathrm{d}$ throughout the upper frequency range, whereas the droop-based variants roll off considerably faster and thereby shift a larger share of the disturbance than intended onto the output capacitance. Around $f_\mathrm{c}$, the I--V $\mathcal{H}_\infty$ and VDCM algorithms enter the disturbance-amplifying region only marginally, by $1.1$ and $0.2\,\mathrm{dB}$, respectively, in contrast to up to $14\,\mathrm{dB}$ for the I--V droop variant. The comparatively contained excursions of the VDCM originate from its voltage control loop, whose integral action restores the bus voltage to its pre-disturbance value, so that the converter acts as a voltage regulator rather than as a droop source: its OII vanishes at DC, as seen in Fig.~\ref{fig:index_plots}(a), and neither steady-state droop nor droop-based load sharing is provided. Also, the emulated inertia yields a pronounced inductive response whose OII phase approaches $+90^\circ$. In summary, the OII and CFI trajectories show that the proposed I--V $\mathcal{H}_\infty$ controller exhibits the desired grid-forming behavior over the widest frequency range among the algorithms considered.

\section{Experimental Results} \label{sec:results_exp}

\subsection{System Setup}
\begin{figure}[t!]
	\centering
	\includegraphics[width=\linewidth]{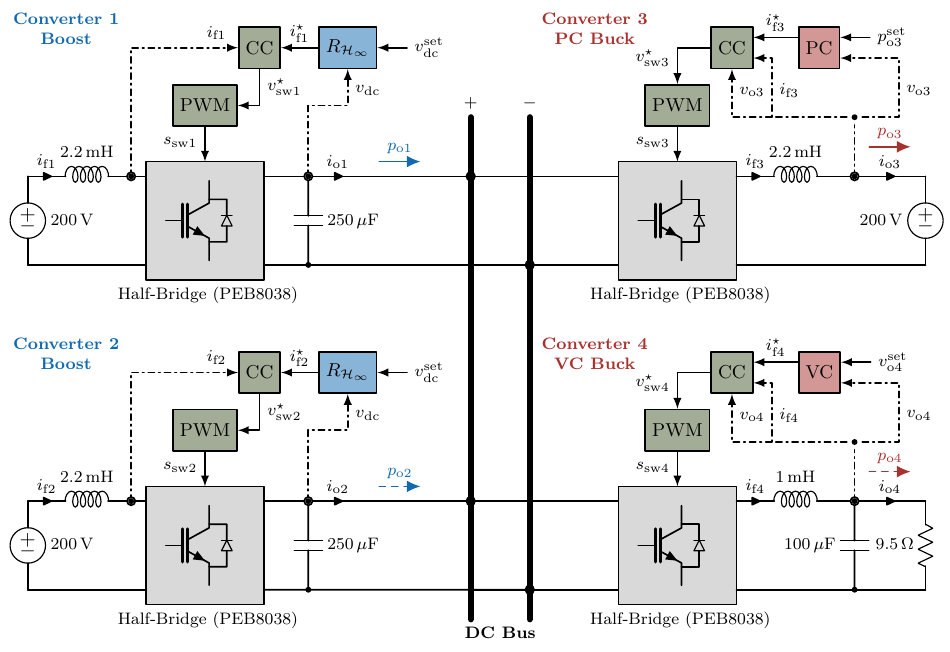}
    \caption{DC microgrid experimental circuit with two boost-type source converters (left) supplying two buck-type load converters (right). The source converters are controlled using the novel I--V $\mathcal{H}_\infty$ control algorithm, while the load converters are controlled using an output Voltage Controller~(VC) and Power Controller~(PC) for the bottom and top unit, respectively.}
	\label{fig:microgrid}
\end{figure}

The proposed I--V $\mathcal{H}_\infty$ controller is validated on a hardware realization of the four-converter DC microgrid depicted in Fig.~\ref{fig:microgrid}, whose parameters are summarized in Table~\ref{tab:parameters}. Two boost converters fed from \SI{200}{\volt} laboratory sources form and stabilize the \SI{350}{\volt} DC bus, and are both controlled by the I--V $\mathcal{H}_\infty$ algorithm. The two buck converters act as loads, with the upper power-controlled unit delivering into a \SI{200}{\volt} laboratory source and the lower voltage-controlled unit regulating \SI{100}{\volt} across a \SI{9.5}{\ohm} resistor. Each half-bridge is realized by an Imperix \mbox{PEB\,8038} module and all control algorithms are executed on a single Imperix \mbox{B-Box\,4} controller at $f_\mathrm{s}=\SI{20}{\kilo\hertz}$, resulting in the total loop delay $T_\mathrm{d}=1.25/f_\mathrm{s}$ listed in Table~\ref{tab:parameters}. The laboratory supplies are triple-output EA~\mbox{PSB\,20920-40}~(Tektronix).

The I--V $\mathcal{H}_\infty$ controller presented in Fig.~\ref{fig:Hinf_K} is discretized using the Tustin transform at $f_\mathrm{s}$ and implemented without any further retuning. At the initial operating point, each boost converter supplies approximately \SI{1}{\kilo\watt} to a total load of \SI{2}{\kilo\watt}, and the DC bus voltage stabilizes at \SI{349.7}{\volt}. A constant feed-forward of \SI{5}{\ampere} is added to the inductor-current reference of each boost unit, such that the droop action only has to supply the remaining current. All waveforms reported below are acquired once per switching period over \SI{200}{\milli\second} windows, and time is expressed relative to the disturbance instant.

\begin{figure}[b!]
	\centering
	\includegraphics[width=1.0\linewidth]{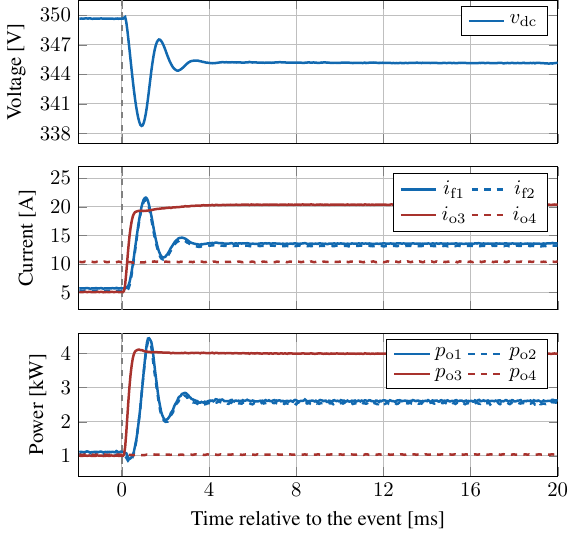}
    \caption{Recorded I--V $\mathcal{H}_\infty$ response of the two source converters to the load increase of the power-controlled buck converter, where the subplots show the DC bus voltage $v_\mathrm{dc}$ (top), the source inductor currents $i_\mathrm{f}$ together with the load currents $i_\mathrm{o}$ (middle), and the output powers $p_\mathrm{o}$ of all four units (bottom).}
	\label{fig:exp_hinf_step}
\end{figure}

\subsection{Load Increase}
Fig.~\ref{fig:exp_hinf_step} shows the response of the two I--V $\mathcal{H}_\infty$-controlled source converters to a load increase, applied as a power step of the power-controlled buck from \SI{1}{\kilo\watt} to \SI{4}{\kilo\watt}. As the load converter completes the step within \SI{0.5}{\milli\second}, the disturbance is seen by the sources as an abrupt output current increase. Following the disturbance, the bus voltage dips to \SI{338.8}{\volt}, i.e., \SI{10.9}{\volt} below the pre-event level, reaching its nadir after \SI{0.9}{\milli\second}. It is subsequently recovered to the droop level of \SI{345.1}{\volt} within \SI{2.75}{\milli\second}, exhibiting a single overshoot of \SI{2.4}{\volt}, which is the time-domain manifestation of the output impedance shaped by the $\mathcal{H}_\infty$ synthesis. Furthermore, the steady-state deviation of \SI{4.53}{\volt} is provided by an additional \SI{7.86}{\ampere} of inductor current at each of the boost converters, i.e., \SI{4.55}{\ampere} referred to the output current of the converters, so that the measured droop gain of \SI{0.995}{\ohm} reproduces the design value $K_\mathrm{d}=\SI{1}{\ohm}$ to within $1\%$.

The recorded currents and powers further illustrate how the voltage response is obtained. As can be seen from the figure, the load current $i_\mathrm{o3}$ steps from \SI{5.1}{\ampere} to \SI{20.4}{\ampere}, so that its power settles at \SI{4}{\kilo\watt}, while the source inductor currents rise from \SI{5.7}{\ampere} to \SI{13.6}{\ampere} and peak at \SI{21.6}{\ampere} in the process. This response is the time-domain counterpart of the impedance analysis of Section~\ref{sec:oii_analysis}. Specifically, a transient excursion appears, namely the oscillations at approximately \SI{600}{\hertz} decaying by about 70\% between successive extrema. This oscillation frequency lies close to the frequency of the \SI{2.5}{\decibel} magnitude overshoot of the OII in Fig.~\ref{fig:index_plots}(a), so that the ringing is the signature of the residual deviation from the desired first-order curve, which is free of resonance and thus of overshoot. Hence, the fast response comes at the expense of this transient excursion, since, with the limited DC-link capacitance, the desired roll-off cannot be achieved under the passivity constraint without a residual overshoot. Although the current limits are not reached in the present experiments, the associated current excursions should be considered when selecting the weighting functions to balance tracking performance and transient speed against reduced current excursions.

\begin{figure}[t!]
	\centering
    \includegraphics[width=1.0\linewidth]{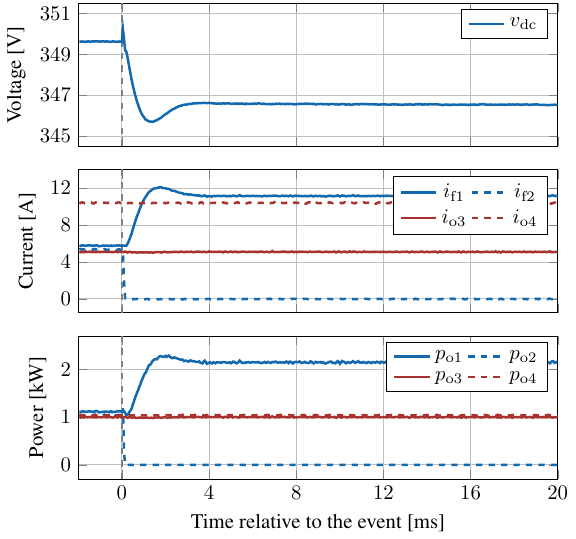}
    \caption{Recorded I--V $\mathcal{H}_\infty$ response to the loss of a grid-forming unit, realized by blocking the PWM of the bottom source converter, showing, from top to bottom, the DC bus voltage $v_\mathrm{dc}$, the source inductor currents $i_\mathrm{f}$ together with the load currents $i_\mathrm{o}$, and the output powers $p_\mathrm{o}$ of all four units.}
	\label{fig:exp_hinf_trip}
\end{figure}

\subsection{Loss of a Grid-Forming Unit}
Fig.~\ref{fig:exp_hinf_trip} illustrates a more severe event, namely the sudden disconnection of the bottom source converter through PWM blocking at the initial dispatch. Its inductor current collapses to zero within a few control periods, such that the upper surviving unit must instantly absorb the entire load, and its power consequently increases from approximately \SI{1}{\kilo\watt} to~\SI{2}{\kilo\watt}. 

Despite the abrupt loss of half the generation capacity, the bus voltage sags by only \SI{3.9}{\volt} to \SI{345.7}{\volt} and returns to the new droop point of \SI{346.5}{\volt} within \SI{1.95}{\milli\second}, without any overshoot. Additionally, the static deviation of \SI{3.09}{\volt} is again matched by \SI{5.39}{\ampere} of additional inductor current, i.e., a bus-referred droop gain of \SI{0.994}{\ohm}. In contrast to the load increase, the transient is here carried entirely by a redistribution of an unchanged total load, and the inductor current of the surviving unit rises smoothly from \SI{5.8}{\ampere} to \SI{11.2}{\ampere} with an excursion of only $9\%$. The two load converters ride through the event essentially unaffected, with their power remaining within $2\%$ of the pre-event value, so that the disturbance is contained by the forming unit.

\section{Conclusion} \label{sec:concl}
This paper proposed a passivity-constrained $\mathcal{H}_\infty$ synthesis methodology for an I--V control architecture, which directly shapes the converter output impedance and current-forming behavior toward their desired grid-forming profiles while enforcing passivity. The $\mathcal{H}_\infty$ controller is constrained to a fixed-order transfer function, so that the resulting scheme remains implementable and interpretable. As the DC-link capacitance is a predefined parameter, the synthesis yields the behavior closest to the desired grid-forming one for the given capacitance.

Benchmarked against representative droop-based and virtual DC machine control schemes from the literature, the resulting I--V $\mathcal{H}_\infty$ controller tracked the desired output impedance and current-forming indices most closely over the relevant frequency range, while exhibiting confined disturbance-amplifying operation. Furthermore, the experimental validation on a four-converter DC microgrid verified the algorithm's practical feasibility and low commissioning effort, owing to the simple structure of the I--V $\mathcal{H}_\infty$ controller and its reliance on a single DC-bus voltage measurement.

Future work will focus on extending the proposed $\mathcal{H}_\infty$ synthesis methodology to load-converter control schemes and developing frequency-selective control designs for the decentralized control of multiple resources in DC microgrids.

\bibliographystyle{IEEEtran}
\bibliography{bibliography}

@ARTICLE{Dragicevic2016,
  author={Dragičević, Tomislav and Lu, Xiaonan and Vasquez, Juan C. and Guerrero, Josep M.},
  journal={IEEE Trans. Power Electron.},
  title={{DC} Microgrids—{Part II}: A Review of Power Architectures, Applications, and Standardization Issues},
  year={2016},
  volume={31},
  number={5},
  pages={3528--3549},
  doi={10.1109/TPEL.2015.2464277}
}

@article{Dragicevic2016a,
  author  = {T. Dragi{\v{c}}evi{\'c} and X. Lu and J. C. Vasquez and J. M. Guerrero},
  title   = {{DC} Microgrids—{Part I}: A Review of Control Strategies and Stabilization Techniques},
  journal = {IEEE Trans. Power Electron.},
  year    = {2016},
  volume  = {31},
  number  = {7},
  pages   = {4876--4891},
  doi={10.1109/TPEL.2015.2478859}
}

@misc{Austermann2024,
  author       = {{Open DC Alliance (ODCA)}},
  title        = {{DC-INDUSTRIE2 System Concept}},
  howpublished = {J.~Austermann and H.~Stammberger, Eds.},
  year         = {2024},
  month        = feb,
  note = {{V}ersion~4},
}

@ARTICLE{Gao2017,
  author={Gao, Fei and Bozhko, Serhiy and Costabeber, Alessandro and Patel, Chintan and Wheeler, Pat and Hill, Christopher Ian and Asher, Greg},
  journal={IEEE Trans. Power Electron.},
  title={Comparative Stability Analysis of Droop Control Approaches in Voltage-Source-Converter-Based {DC} Microgrids},
  year={2017},
  volume={32},
  number={3},
  pages={2395--2415},
  doi={10.1109/TPEL.2016.2567780}
}

@INPROCEEDINGS{Jin2017a,
  author={Jin, Zheming and Meng, Lexuan and Guerrero, Josep M.},
  booktitle={IEEE Second International Conference on DC Microgrids (ICDCM)},
  title={Comparative admittance-based analysis for different droop control approaches in {DC} microgrids},
  year={2017},
  pages={515--522},
  doi={10.1109/ICDCM.2017.8001095}
}

@ARTICLE{Wang2018,
  author={Wang, Haojie and Han, Minxiao and Han, Renke and Guerrero, Josep M. and Vasquez, Juan C.},
  journal={IEEE Trans. Power Electron.},
  title={A Decentralized Current-Sharing Controller Endows Fast Transient Response to Parallel {DC}–{DC} Converters},
  year={2018},
  volume={33},
  number={5},
  pages={4362--4372},
  doi={10.1109/TPEL.2017.2714342}
}

@ARTICLE{Zhu2020,
  author={Zhu, Xiaorong and Meng, Fanqi and Xie, Zhiyun and Yue, Ying},
  journal={IEEE Trans. Energy Convers.},
  title={An Inertia and Damping Control Method of {DC}–{DC} Converter in {DC} Microgrids},
  year={2020},
  volume={35},
  number={2},
  pages={799--807},
  doi={10.1109/TEC.2019.2952717}
}

@ARTICLE{Deng2024,
  author={Deng, Fei and Zhang, Lei and Xiao, Ziheng and Yao, Zhigang and Tang, Yi},
  journal={IEEE Trans. Power Electron.},
  title={Inertia Emulation in Droop-Based {DC} Microgrids With Equivalent Converter Impedance Reshaping},
  year={2024},
  volume={39},
  number={11},
  pages={15206--15216},
  doi={10.1109/TPEL.2024.3437767}
}

@ARTICLE{Liu2020,
  author={Liu, Guangyuan and Mattavelli, Paolo and Saggini, Stefano},
  journal={IEEE Trans. Power Electron.},
  title={Resistive-Capacitive Output Impedance Shaping for Droop-Controlled Converters in {DC} Microgrids With Reduced Output Capacitance},
  year={2020},
  volume={35},
  number={6},
  pages={6501--6511},
  doi={10.1109/TPEL.2019.2951729}
}

@article{Na2022,
  author    = {Na Zhi and Xu Ming and Youguo Ding and Liang Du and Hui Zhang},
  title     = {Power-Loop-Free Virtual {DC} Machine Control With Differential Compensation},
  journal   = {IEEE Trans. Ind. Appl.},
  volume    = {58},
  number    = {1},
  pages     = {413--422},
  year      = {2022},
  doi={10.1109/TIA.2021.3119512}
}

@article{Samanta2018,
  author    = {Soumya Samanta and Jyoti Prakash Mishra and Binoy Krishna Roy},
  title     = {Virtual {DC} machine: An inertia emulation and control technique for a bidirectional {DC}–{DC} converter in a {DC} microgrid},
  journal   = {IET Electr. Power Appl.},
  volume    = {12},
  number    = {6},
  pages     = {874--884},
  year      = {2018},
  doi       = {10.1049/iet-epa.2017.0770},
  publisher = {Institution of Engineering and Technology}
}

@inproceedings{Tan2016,
  author    = {Shucheng Tan and Ge Dong and Hui Zhang and Na Zhi and Xi Xiao},
  title     = {Virtual {DC} Machine Control Strategy of Energy Storage Converter in {DC} Microgrid},
  booktitle = {IEEE Electrical Power and Energy Conference (EPEC)},
  year      = {2016},
  doi       = {10.1109/EPEC.2016.7771749},
}

@article{Lin2021,
  author    = {Gang Lin and Junjie Ma and Yong Li and Christian Rehtanz and Jiayan Liu and Ziya Wang and Pengcheng Wang and Fengjian She},
  title     = {A Virtual Inertia and Damping Control to Suppress Voltage Oscillation in Islanded {DC} Microgrid},
  journal   = {IEEE Trans. Energy Convers.},
  volume    = {36},
  number    = {3},
  pages     = {2134--2144},
  year      = {2021},
  doi       = {10.1109/TEC.2020.3039364},
  publisher = {IEEE}
}

@INPROCEEDINGS{Jin2018,
  author={Jin, Zheming and Guerrero, Josep M. and Li, Mingshen},
  booktitle={IEEE Energy Conversion Congress and Exposition (ECCE)},
  title={An Alternative Realization of Droop Control and Virtual Impedance for Paralleled Converters in {DC} Microgrid},
  year={2018},
  pages={3765--3770},
  doi={10.1109/ECCE.2018.8557628}
}

@INPROCEEDINGS{Jin2017b,
  author={Jin, Zheming and Meng, Lexuan and Han, Renke and Guerrero, Josep M. and Vasquez, Juan C.},
  booktitle={IEEE Energy Conversion Congress and Exposition (ECCE)},
  title={Admittance-type {RC}-mode Droop Control to Introduce Virtual Inertia in {DC} Microgrids},
  year={2017},
  pages={4107--4112},
  doi={10.1109/ECCE.2017.8096714}
}

@article{Justo2013,
title = {{AC}-microgrids versus {DC}-microgrids with distributed energy resources: A review},
journal = {Renew. Sustain. Energy Rev.},
volume = {24},
pages = {387--405},
year = {2013},
issn = {1364-0321},
doi = {10.1016/j.rser.2013.03.067},
author = {Jackson John Justo and Francis Mwasilu and Ju Lee and Jin-Woo Jung},
}

@INPROCEEDINGS{Lazarevic2025,
  author={Lazarevi{\'c}, Vladan {\v{Z}}. and Schweizer, Mario and Pejovski, Dejan and Antoniazzi, Antonello},
  booktitle={IEEE Seventh International Conference on DC Microgrids (ICDCM)},
  title={Ensuring Stability in {DC} Microgrids Through Application of the Passivity Principle},
  year={2025},
  doi={10.1109/ICDCM63994.2025.11144700}
}

@ARTICLE{Harnefors1998,
  author={Harnefors, L. and Nee, H.-P.},
  journal={IEEE Trans. Ind. Appl.},
  title={Model-based current control of {AC} machines using the internal model control method},
  year={1998},
  volume={34},
  number={1},
  pages={133--141},
  doi={10.1109/28.658735}
}

@INPROCEEDINGS{Krajacic2026a,
  author={Krajacic, Jovan and Stanojev, Ognjen and Schweizer, Mario and Karaca, Orcun and Hug, Gabriela and Lazarević, Vladan},
  booktitle={IEEE Eighth International Conference on DC Microgrids (ICDCM)},
  title={Grid-Forming Characterization in {DC} Microgrids},
  year={2026},
  doi={10.1109/ICDCM67837.2026.11594507}
}

@INPROCEEDINGS{Krajacic2026b,
  author={Krajacic, Jovan and Stanojev, Ognjen and Schweizer, Mario and Karaca, Orcun and Hug, Gabriela and Lazarević, Vladan},
  booktitle={IEEE Eighth International Conference on DC Microgrids (ICDCM)},
  title={Exploring Converter Control Duality in Microgrids: {AC} Grid-Forming vs {DC} Droop Control},
  year={2026},
  doi={10.1109/ICDCM67837.2026.11594561}
}

@ARTICLE{Apkarian2006,
  author={Apkarian, P. and Noll, D.},
  journal={IEEE Trans. Autom. Control},
  title={Nonsmooth {$H_\infty$} Synthesis},
  year={2006},
  volume={51},
  number={1},
  pages={71--86},
  doi={10.1109/TAC.2005.860290}
}

@INPROCEEDINGS{Apkarian2014,
  author={Apkarian, Pierre and Gahinet, Pascal and Buhr, Craig},
  booktitle={European Control Conference (ECC)},
  title={Multi-model, multi-objective tuning of fixed-structure controllers},
  year={2014},
  pages={856--861},
  doi={10.1109/ECC.2014.6862200}
}

@article{Bruinsma1990,
title = {A fast algorithm to compute the {$H_\infty$}-norm of a transfer function matrix},
journal = {Syst. Control Lett.},
volume = {14},
number = {4},
pages = {287--293},
year = {1990},
issn = {0167-6911},
doi = {10.1016/0167-6911(90)90049-Z},
author = {N.A. Bruinsma and M. Steinbuch},
}

@article{DArco2015,
title = {A Virtual Synchronous Machine implementation for distributed control of power converters in Smart Grids},
author={D'Arco, Salvatore and Suul, Jon Are and Fosso, Olav B},
journal = {Electr. Power Syst. Res.},
volume = {122},
pages = {180--197},
year = {2015},
issn = {0378-7796},
  doi={10.1016/j.epsr.2015.01.001}
}

@techreport{GEVernova2024,
  author      = {{GE Vernova Power Conversion}},
  title       = {{DC Microgrids}},
  institution = {GE Vernova},
  number      = {GEA25524},
  year        = {2024},
  month       = jul,
  type        = {White Paper}
}

@book{Zhou1996,
  author    = {Zhou, Kemin and Doyle, John C. and Glover, Keith},
  title     = {Robust and Optimal Control},
  publisher = {Prentice Hall},
  address   = {Upper Saddle River, NJ, USA},
  year      = {1996}
}

@book{Khalil2002,
  author    = {Hassan K. Khalil},
  title     = {Nonlinear Systems},
  edition   = {3rd},
  publisher = {Prentice Hall},
  year      = {2002}
}

@ARTICLE{Chen2022,
  author={Chen, Meng and Zhou, Dao and Tayyebi, Ali and Prieto-Araujo, Eduardo and Dörfler, Florian and Blaabjerg, Frede},
  journal={IEEE Trans. Smart Grid},
  title={Generalized Multivariable Grid-Forming Control Design for Power Converters},
  year={2022},
  volume={13},
  number={4},
  pages={2873--2885},
  doi={10.1109/TSG.2022.3161608}
}

@ARTICLE{Haberle2022,
  author={Häberle, Verena and Fisher, Michael W. and Prieto-Araujo, Eduardo and Dörfler, Florian},
  journal={IEEE Trans. Power Syst.},
  title={Control Design of Dynamic Virtual Power Plants: An Adaptive Divide-and-Conquer Approach},
  year={2022},
  volume={37},
  number={5},
  pages={4040--4053},
  doi={10.1109/TPWRS.2021.3139775}
}

@ARTICLE{Yang2011,
  author={Yang, Shuitao and Lei, Qin and Peng, Fang Z. and Qian, Zhaoming},
  journal={IEEE Trans. Ind. Electron.},
  title={A Robust Control Scheme for Grid-Connected Voltage-Source Inverters},
  year={2011},
  volume={58},
  number={1},
  pages={202--212},
  doi={10.1109/TIE.2010.2045998}
}

@ARTICLE{Kammer2019,
  author={Kammer, Christoph and D’Arco, Salvatore and Endegnanew, Atsede Gualu and Karimi, Alireza},
  journal={IEEE Trans. Power Electron.},
  title={Convex Optimization-Based Control Design for Parallel Grid-Connected Inverters},
  year={2019},
  volume={34},
  number={7},
  pages={6048--6061},
  doi={10.1109/TPEL.2018.2881196}
}

@ARTICLE{Huang2020,
  author={Huang, Linbin and Xin, Huanhai and Dörfler, Florian},
  journal={IEEE Trans. Smart Grid},
  title={{$H_\infty$}-Control of Grid-Connected Converters: Design, Objectives and Decentralized Stability Certificates},
  year={2020},
  volume={11},
  number={5},
  pages={3805--3816},
  doi={10.1109/TSG.2020.2984946}
}

@misc{Stanojev2026,
      title={Improved Droop Control in {DC} Microgrids via Voltage-Locked Loop Synchronization},
      author={Ognjen Stanojev and Jovan Krajacic and Enea Bianda and Orcun Karaca and Mario Schweizer},
      year={2026},
      eprint={2609.21913},
      archivePrefix={arXiv},
      primaryClass={eess.SY},
      url={https://arxiv.org/abs/2609.21913},
}

@misc{currentos2025,
  author       = {{Current/OS Standard}},
  title        = {System Reference Document},
  year         = {2025},
}

@ARTICLE{Serrano2020,
  author={Serrano-Delgado, Javier and Mattavelli, Paolo and Cobreces, Santiago and Abedini, Hossein and Rizo, Mario and Buso, Simone and Bueno, Emilio Jose},
  journal={IEEE Access},
  title={Output Capacitance Minimization for Converters in {DC} Microgrids via Multi-Objective Tuning of Droop-Based Controllers},
  year={2020},
  volume={8},
  pages={222700--222710},
  doi={10.1109/ACCESS.2020.3044171}
}

@INPROCEEDINGS{Tripathy2025,
  author={Tripathy, Rudrakant and Maulik, Avirup and Sharma, Ruchi and Kamal, Shyam},
  booktitle={IEEE 5th International Conference on Sustainable Energy and Future Electric Transportation (SEFET)},
  title={Virtual Inertia Implementation with Robust Voltage Controller for a Solar-Battery {DC} Microgrid},
  year={2025},
  doi={10.1109/SEFET65155.2025.11255304}
}

@INPROCEEDINGS{Ren2021,
  author={Ren, Haoyan and Ma, Lei},
  booktitle={IEEE International Conference on Power, Intelligent Computing and Systems (ICPICS)},
  title={{H-infinity} Mixed Sensitivity Control of Dual Active Bridge {DC-DC} Converter},
  year={2021},
  pages={16--20},
  doi={10.1109/ICPICS52425.2021.9524165}
}

@ARTICLE{Liu2024,
  author={Liu, Ziyong and Ebrahimi, Javad and Gholizadeh Narm, Hossein and Eren, Suzan},
  journal={IEEE J. Emerg. Sel. Topics Power Electron.},
  title={Smart Search Implemented {H-Infinity} Control Design for {DAB} Converter in {DC} Microgrid},
  year={2024},
  volume={12},
  number={2},
  pages={1906--1920},
  doi={10.1109/JESTPE.2024.3351130}
}

@INPROCEEDINGS{Liu2025,
  author={Liu, Ziyong and Ebrahimi Jouzdani, Javad and Eren, Suzan},
  booktitle={IEEE 34th International Symposium on Industrial Electronics (ISIE)},
  title={Novel {H-Infinity} Controller for {DAB} Converter under Constant Current Load},
  year={2025},
  doi={10.1109/ISIE62713.2025.11124786}
}

@INPROCEEDINGS{Liu2026,
  author={Liu, Ziyong and Ebrahimi, Javad and Eren, Suzan},
  booktitle={IEEE Applied Power Electronics Conference and Exposition (APEC)},
  title={Robust Control Design and Implementation for Dual Active Bridge Converter with Constant Power Load Under Parameter Uncertainty},
  year={2026},
  pages={2885--2892},
  doi={10.1109/APEC51134.2026.11516832}
}

@misc{MATLAB,
  author    = {{The MathWorks Inc.}},
  title     = {{MATLAB} version: R2024b},
  year      = {2024},
  publisher = {The MathWorks Inc.},
  address   = {Natick, Massachusetts, United States},
  url       = {https://www.mathworks.com}
}

@ARTICLE{Jithin2023,
  author={Jithin, Kolampurath and Haridev, Puthan Purayil and Mayadevi, Nanappan and Harikumar, Raveendran Pillai and Mini, Valiyakulam Prabhakaran},
  journal={J. Mod. Power Syst. Clean Energy},
  title={A Review on Challenges in {DC} Microgrid Planning and Implementation},
  year={2023},
  volume={11},
  number={4},
  pages={1375--1395},
  doi={10.35833/MPCE.2022.000053}
}

\end{document}